\documentclass[12pt]{article}

\usepackage{amsmath, amssymb, amsfonts}
\usepackage{graphicx}

\usepackage{scicite}

\renewcommand{\phi}{ \varphi }

\newcommand{\degC}[1]{{$^{\rm\circ}$}}

\usepackage{times}

\newenvironment{sciabstract}{%
	\begin{quote} \bf}
	{\end{quote}}

\title{Silicon Flexes Muscles:\\ Giant Electrochemical Actuation in a Nanoporous Silicon-Polypyrrole Hybrid Material}

\author
{Manuel Brinker,${}^{1}$ Guido Dittrich,${}^{1}$ Claudia Richert,${}^{2}$ Pirmin Lakner,${}^{3,4}$\\ Tobias Krekeler,${}^{5}$ Thomas F. Keller,${}^{3,4}$ Norbert Huber${}^{2}$ and Patrick Huber${}^{1,3,6\ast}$\\
	\\
	\normalsize{${}^{1}$Institute of Materials Physics and X-ray Analytics ,} \\ \normalsize{Hamburg University of Technology, 21073 Hamburg, Germany,}\\
	\normalsize{${}^{2}$Institute of Materials Research, Materials Mechanics,} \\ \normalsize{Helmholtz-Zentrum Geesthacht, Geesthacht 21502, Germany,}\\
	\normalsize{${}^{3}$Centre for X-ray and Nano Science CXNS,}\\\normalsize{Deutsches Elektronen-Synchrotron DESY, Notkestra\ss e 85, 22607 Hamburg, Germany,}\\
	\normalsize{${}^{4}$Physics Department,}\\ \normalsize{University of Hamburg, 20355 Hamburg, Germany,}\\
	\normalsize{${}^{5}$Electron Microscopy Unit,}\\ \normalsize{Hamburg University of Technology, 21073 Hamburg, Germany}\\
	\normalsize{${}^{6}$Centre for Hybrid Nanostructures CHyN,}\\ \normalsize{University of Hamburg, 22607 Hamburg, Germany}\\
	\\
	\normalsize{${}^\ast$To whom correspondence should be addressed; E-mail:  patrick.huber@tuhh.de.}
}

\date{}

\begin{document} 
	
	
	\baselineskip24pt
	
	
	\maketitle

	\begin{sciabstract}
		The absence of piezoelectricity in silicon makes direct electro-mechanical applications of this mainstream semiconductor impossible. Integrated electrical control of the silicon mechanics, however, would open up new perspectives for on-chip actuorics. Here, we combine wafer-scale nanoporosity in single-crystalline silicon with polymerization of an artificial muscle material inside pore space to synthesize a composite that shows macroscopic electrostrain in aqueous electrolyte. The voltage-strain coupling is 3 orders of magnitude larger than the best-performing ceramics in terms of piezoelectric actuation. We trace this huge electroactuation to the concerted action of 100 billions of nanopores per square centimetre cross-section and to potential-dependent pressures of up to 150 atmospheres at the single-pore scale. The exceptionally small operation voltages (0.4\,-\,0.9\,V) along with the sustainable and biocompatible base materials make this hybrid promising for bio-actuator applications.
	\end{sciabstract}
	\textbf{Summary:} A silicon-polymer hybrid shows large, reversible electrostrain at remarkably small operation voltages in aqueous electrolyte.
	\section*{Introduction}
	\noindent An electrochemical change in the oxidation state of polypyrrole (PPy) can increase or decrease the number of delocalized charges in its polymer backbone \cite{Ma2013}. Immersed in an electrolyte this is also accompanied by a reversible counter-ion uptake or expulsion and thus with a marcroscopic contraction or swelling under electrical potential control, making PPy one of the most employed artificial muscle materials \cite{Jager2000,Madden2002,Smela2003,Ma2013,Balint2014}.\\
	%In this work we demonstrate PPy-filling of a detached porous silicon layer with an aspect ratio of the pore-length to -diameter of $11800$. We show that the resulting nanoporous monolithic material exhibits exceptional actuation properties.\\	
	\noindent 
	Here, we combine this actuator polymer with the 3D scaffold structure of nanoporous silicon \cite{Lin1997,Sailor2011,Canham2015} to design, similarly as found in many multiscale biological composites in nature \cite{Eder2018}, a material with embedded electo-chemical actuation that consists of a few light and abundant elemental constituents (i.e. H, C, N, O, Si, Cl).\\
	\section*{Results}
	\begin{figure}[]
		\centering
		\includegraphics[width=0.5\textwidth]{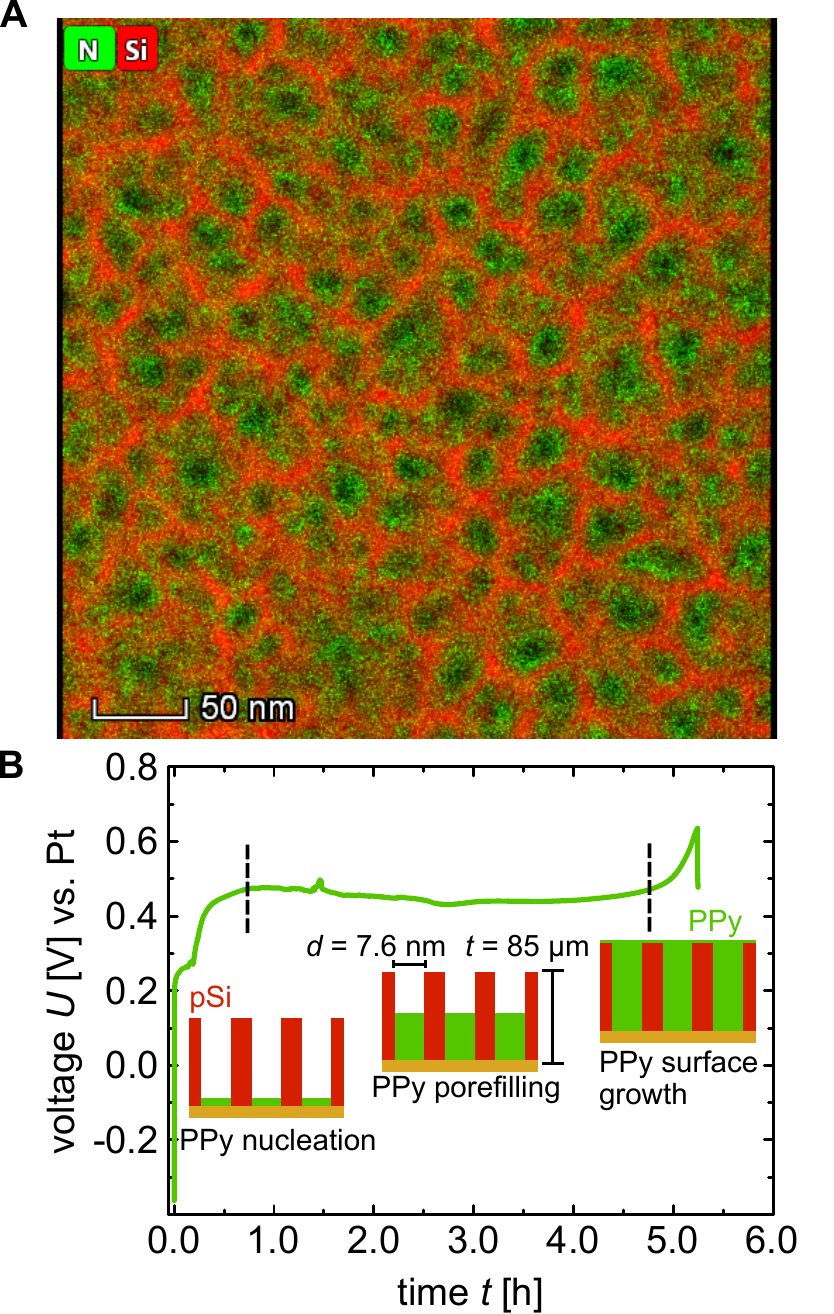}
		\caption{\textbf{Synthesis of a nanoporous polypyrrole-silicon material.} (\textbf{A}) High-angle annular dark-field scanning transmission electron micrograph top-view on a nanoporous silicon membrane filled by electropolymerisation with pyrrole. The green and red color code indicates the N and Si concentration resulting from energy-dispersive x-ray detection (EDX) measurements, respectively. (\textbf{B}) Voltage-time recording during galvanostatic electropolymerisation of pyrrole in nanoporous silicon, with mean pore diameter $d$ and thickness $t$.  Characteristic regimes are indicated and discussed in the text.}
		\label{Fig_Synthesis}
	\end{figure}
	In a first step the porous silicon membrane is prepared in an electrochemical etching process of doped silicon in hydrofluoric acid. The resulting pores are characteristically straight and perpendicular to the silicon surface. The analysis of a nitrogen sorption measurement yields a mean pore diameter of $d =7.2\,\mathrm{nm}$ and a porosity of $\Phi = 50\,\%$. This corresponds to 170 billion channels per square-centimeter. Scanning electron microscopy profiles give a homogeneous sample thickness of $t=85\pm 1\,\mathrm{ \mu  m}$.
	The pSi membrane is then filled with PPy by electropolymerisation of pyrrole monomers. The graph in Fig.~1B shows the time-dependent voltage during the galvanostatic polymerisation process. The curve exhibits characteristic transient regimes. Initially the voltage shows an increase from its open circuit potential of around $-0.4\,\mathrm{V}$ to approximately $0.5\,\mathrm{V}$. This potential transition is attributed to the polymer nucleation at the bottom of the pore and a partial oxidation of the porous silicon. The ensuing constant voltage plateau is then related to a constant deposition of polypyrrole inside the pores. The voltage increase at $5$ hours is associated with the termination of pore filling and the onset of polymerisation of pyrrole on the membrane surface. The geometrical constraint of the highly asymmetrical pores benefits a chainlike polymer growth and inhibits a branching of the polymer, which leads to a lower electrical resistance \cite{Schultze1995}. Conversely, a polymer branching is more likely to appear during the unrestricted polymerisation on the surface leading to a higher electrical resistance. Thus, a transition to a higher voltage indicates the polymerisation of bulk PPy on the surface. To prevent this layer formation, the current is switched off when the voltage starts to increase towards a higher voltage, as it is the case at $5$ hours. A transmission electron micrograph (TEM) with an overlayed energy dispersive X-ray spectroscopy (EDX) signal of the resulting membrane materials indicates a homogeneous PPy filling of the random pSi honeycomb structure, see Fig.~1A. \\
	To characterize the actuorics of the resulting hybrid material dilatometry measurements in an in-situ electrochemical setup are performed. The sample is immersed in perchloric acid ($\mathrm{HClO}_4$ with a concentration $1\,\mathrm{mol}/\mathrm{l}$) and positioned, so that the pores are pointing in a horizontal direction, see Fig.~2A and S1C in the Supplementary Material. The quartz probe of the dilatometer is positioned on the top of the sample in order to measure the sample-length $l$. The sample length $l_0$ that is thus in contact with the perchloric acid is $l_0=0.626\pm 0.005\,\mathrm{mm}$ with a width of the sample of $w=3.49\pm 0.01\,\mathrm{mm}$. The  measurements are performed with a potentiostat in a three electrode setup. All potentials are indicated versus the standard hydrogen potential (SHE).\\
	\iftrue
	\begin{figure*}[]
		\centering
		\includegraphics[width=1\textwidth]{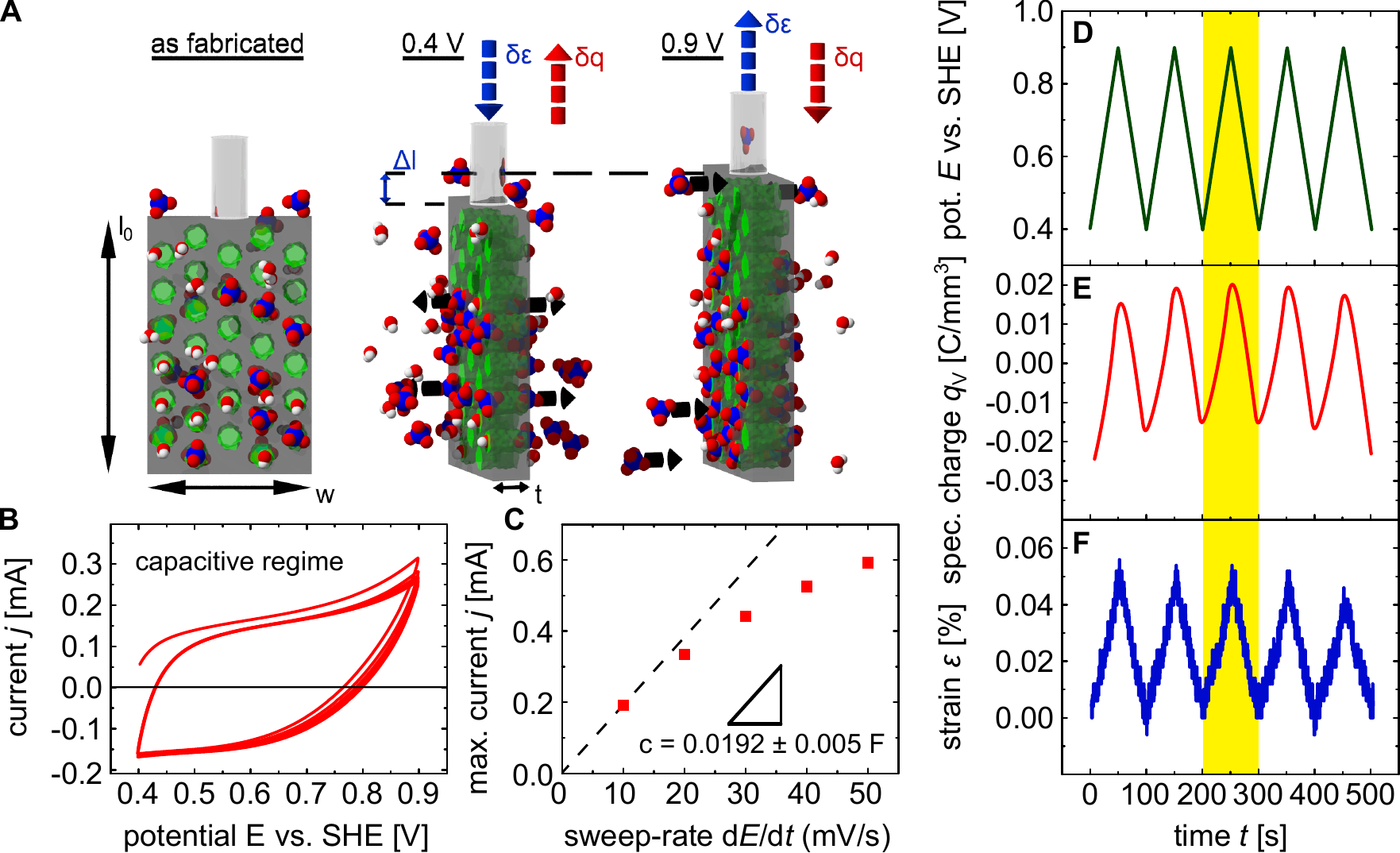}
		\caption{\textbf{Electrochemical actuation experiments.} (\textbf{A}) Schematics of the electroactuation experiments on the pSi membrane (grey) filled with PPy (green) immersed in an aqueous electrolyte ($\mathrm{HClO}_4$ (blue, red) and $\mathrm{H}_2\mathrm{O}$ (red, white) molecules). The dimensions of the as fabricated membrane, on the left, are length $l_0$, width $w$ and thickness $t$. The middle part illustrates the case where a voltage of $0.4\,\mathrm{V}$ is applied and the $\mathrm{ClO}_4^{-}$ anions are expelled from the PPy resulting in the contraction of the sample. Vice versa, in the right part a voltage of $0.9\,\mathrm{V}$ is applied and the anions are incorporated followed by the subsequent expansion of the sample. The change in length is indicated by $\Delta l$. (\textbf{B}) The graph depicts an exemplary cyclic voltammetry of a pSi-PPy membrane in $1\,\mathrm{mol}/\mathrm{l}$  $\mathrm{HClO}_4$ electrolyte. The current $j$ is plotted against the applied potential $E$ measured versus the standard hydrogen potential (SHE). The potential sweep rate is $10\,\mathrm{mV}/\mathrm{s}$. (\textbf{C}) The graph depicts the mean values for the maximal current density of $j$ plotted against varying potential sweep rates $\mathrm{d}E/\mathrm{d}t$ from $10\,\mathrm{mV}/\mathrm{s}$ to $50\,\mathrm{mV}/\mathrm{s}$. The dashed line indicates a linear regression of the data points which yields the capacitance $c^{*}$ as the slope. Depicted on the right are (\textbf{D}) 5 representative potential cycles $E$, (\textbf{E}) the resulting volumetric charge $q_\mathrm{V}$ and (\textbf{F}) the introduced effective strain $\varepsilon$ of the nanoporous membrane.}
		\label{Fig_Actuation}
	\end{figure*}
	Before and during the actuoric dilatometry measurements the electrochemical characteristics of the hybrid system are determined by recording cyclic voltammograms (CV) in the potential range from $0.4\,\mathrm{V}$ to $0.9\,\mathrm{V}$ with a sweep rate of $10\,\mathrm{mV}/\mathrm{s}$, see Fig.~2B. The ordinate shows the current $j$. The pSi-PPy membrane exhibits the characteristics of a capacitive charging of the PPy with $j$ quickly moving towards a constant value and the absence of true oxidation and reduction peaks \cite{Madden2001}. Higher voltages are not applied, since it leads to an overoxidation and partial destruction of PPy \cite{Lewis1997}. Lower voltages could be applied and a true reduction peak should be visible below $0\,\mathrm{V}$. However, an aqueous electrolyte is used which limits the CVs lower sweep potential to values above $0\,\mathrm{V}$, since electrolysis of the water sets in  below \cite{Roschning2019}. But Faradaic current contributions are present in the depicted CV since the currents towards the upper limit and the lower limit are not completely constant, as they ideally should be for pure capacitive behavior. The Faradaic currents can be attributed to an oxidative process that consumes charge. Likely, it is an oxidation of the silicon pore walls. We account for this effect by subtraction of the charge that is consumed in this way.\\
	In Fig.~2B we present the dependence of the maximal current $j_\mathrm{max}$ recorded in the CV, plotted against an increasing scan rate $\dot{E} = \mathrm{d}E / \mathrm{d}t$.
	The non linear course of the relation shows that a diffusion limitation, discussed in more detail below, affects the capacitive charging of the PPy. The limitation is larger for higher sweep rates. The capacitance $c$ is defined as $c = \delta q / \delta E$, where $q $ is the charge and $E$ is the applied potential.
	The linear slope of $j_\mathrm{max}$ plotted against the sweep rate gives then $c = \Delta j_\mathrm{max} / \Delta (\mathrm{d}E / \mathrm{d}t) $. Because of the diffusion limitation the capacitance is determined by a linear regression of the first data point only, $c = 0.0192\pm 0.005\,\mathrm{F}$. Assuming a volume specific capacity of $0.24\,\mathrm{F}/\,\mathrm{mm^3}$, as reported for PPy films doped with the same anion, the PPy confined in the pores has a volume of $V_\mathrm{PPy} = 0.08\,\mathrm{mm^3}$. This corresponds to a reasonable pore filling degree of $86.2\,\%$.\\	
	Recording the sample length change while recording the CVs allows as a detailed characterisation of the electrochemical actuation, see Fig.~2D-F. While $E$ is reversibly changed from $0.4\,\mathrm{V}$ to $0.9\,\mathrm{V}$, the current and thus the transferred charge is recorded. The charge normalized to the sample volume $V_\mathrm{sample}$ is denoted as $q_\mathrm{V}$ and shown in Fig.~2E. The transferred charge clearly coincides with $E$. So, increasing the potential leads to an increased insertion of anions into the PPy and, vice versa, decreasing the potential releases anions from the PPy.
	The resulting length variation normalized to the sample length $\varepsilon = \Delta l / l_0 = (l-l_0)/l_0$ is presented in Fig.~2F. The repetitive change in sample length clearly coincides linearly with $E$ and $q_\mathrm{V}$. So, the incorporation of anions leads to an expansion of the sample while the release of the anions lets the sample contract. The amplitude of the length change shows no signs of a decrease and is exhibiting an excellent reproducibility and thus a repeatable and robust mechanical actuation functionality. The sign of the strain-charge coupling is positive, meaning that sweeping the potential in positive direction leads to an expansion of the sample and vice versa.\\
	\begin{figure*}[]
		\centering
		\includegraphics[width=1\textwidth]{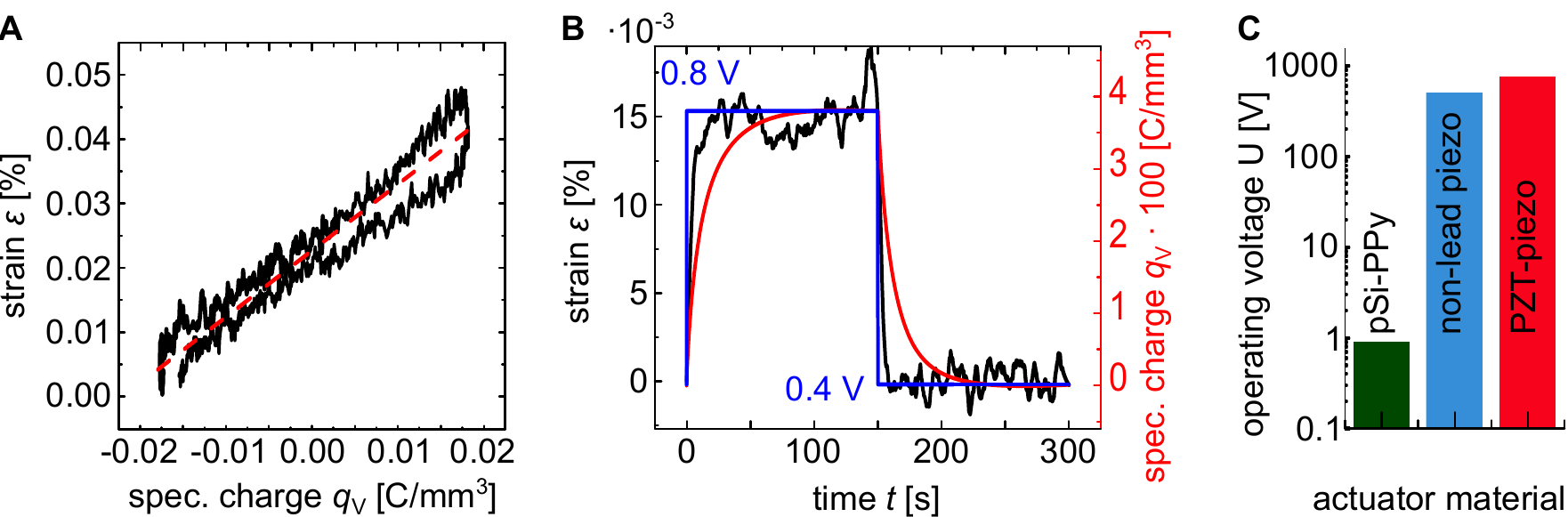}
		\caption{\textbf{Static and dynamic electrochemical performance parameters.} (\textbf{A}) Strain $\varepsilon$ versus deposited volumetric charge $q_\mathrm{V}$. (\textbf{B}) Step-coulombmetry to determine the electroactuation kinetics. The applied potential $E$ is changed in a step-like fashion from $0.4\,\mathrm{V}$ to $0.8\,\mathrm{V}$ and backwards while the thereby incorporated volumetric charge $q_\mathrm{V}$ and the caused strain $\varepsilon$ are measured versus time. (\textbf{C}) Operating voltage $U$ needed to achieve a strain amplitude of $\varepsilon = 0.05\,\%$ for the sample length of $l=0.626\,\mathrm{mm}$ for the pSi-PPy sample in comparison with high-performance piezoelectric lead-free \cite{Liu2009b} and lead-containing (PZT) ceramics \cite{Saito2004}.}
		\label{Fig_Kinetics}
	\end{figure*}
	Additional actuation details can be inferred by averaging the volumetric charge density $q_\mathrm{V}$, normalized to the sample volume, and the strain $\varepsilon$ over the 5 CV cycles, separately for positive- and negative-pointing sweeps. The peak-to-peak amplitude of the relative length variation $\varepsilon$ over the 5 cycles is $0.05\pm 0.002\,\%$, which yields an absolute amplitude of $0.313\pm 0.0013\,\mathrm{\mu m}$. Fig.~3A provides a plot of $\varepsilon$ vs. $q_\mathrm{V}$, both recorded in the associated potential range of Fig.~2C. The relation between the two is highly linear as well in agreement with the robust actuation functionality inferred above. The corresponding linear coupling parameter relating the strain response to the incorporated anions by the volumetric charge \cite{Stenner2016} is $A^{*}= \mathrm{d}\varepsilon/\mathrm{d}q_{\mathrm{V}} = 0.01029 \pm 0.00009\,\mathrm{mm^3}/\mathrm{C}$.\\
	To analyse the actuation kinetics step-coulombmetry is performed. The response to a square potential from $0.4\,\mathrm{V}$ to $0.8\,\mathrm{V}$ is depicted in Fig.~3B. The associated strain $\varepsilon$ saturates at $0.015\,\%$, while the volumetric charge $q_\mathrm{V}$ saturates at $0.04\,\mathrm{C}/\mathrm{mm^3}$. When the potential is decreased to $0.4\,\mathrm{V}$, $\varepsilon$ decreases to  $0\,\%$ and the volumetric charge reaches $0.0\,\mathrm{C}/\mathrm{mm^3}$. Simple exponential functions can be used to fit the course of $\varepsilon$ and $q_\mathrm{V}$ for both the increase and decrease respectively. The time constants $\tau$ yield the speed of the respective processes, i.e. the charging and discharging of $q_\mathrm{V}$ and the respective reaction of $\varepsilon$: $\tau_{q_\mathrm{V},\mathrm{incr}} = 16.52 \pm 0.07\,\mathrm{s}$ for the volumetric charge increase,  $\tau_{\varepsilon,\mathrm{incr}} = 4.26 \pm 0.17\,\mathrm{s}$ for the strain increase, $\tau_{q_\mathrm{V},\mathrm{decr}} = 14.31 \pm 0.07\,\mathrm{s}$ for the volumetric charge decrease and $\tau_{\varepsilon,\mathrm{decr}} = 2.8 \pm 0.1\,\mathrm{s}$ for the strain decrease.
	Interestingly, the strain response is thus faster than the charging and discharging process by almost an order of magnitude. Two effects likely contribute to this observation. Firstly, the PPy reaches, at least partially, its yield limit and plastic deformation of the PPy sets in. Thus the whole sample is not expanding further, even though counter-ions are still incorporated into the polymer, a hypothesis which we address in more detail in the micromechanical analysis below. Secondly, the already mentioned diffusion limitation hinders a faster transfer of the anions to the PPy. An estimation of the anion drift dynamics based on the self-diffusion coefficient $\mathrm{D}= 1.53\cdot 10^{-9}\,\mathrm{m^2 s^{-1}}$ \cite{Heil1995} of $\mathrm{ClO}_{4}^{-}$ ions in PPy, as determined by Molecular Dynamics simulations, \cite{Lopez2004} supports this kinetic limitation. It gives a characteristic diffusion time of $59\,\mathrm{s}$ for the path of $42.5\,\mathrm{\mu m}$ between bulk reservoir and membrane centre, which is on the same order of magnitude determined for $\tau_{q_\mathrm{V},\mathrm{incr}}$ and $\tau_{q_\mathrm{V},\mathrm{decr}}$. It is also interesting to note, that the sample contracts faster by a factor of approximately $2$ than it is expanding. This can be rationalized by the distinct mesoscopic mechanisms upon contraction and expansion. In particular, upon potential reversal the highly strained Si lattice supports the expulsion of the ions.\\ 
	\section*{Discussion}
	For obtaining a further understanding of the electroactuation mechanism of the PPy filled pSi membranes, the micromechanical properties are modelled based on the microstructure as extracted from electron micrograph of the same area of material presented in Fig.~1A. Details on the translation of the greyscale image into a finite element model are given in the methods section \cite{Gibson1982,Huber2018,Richert2018}. 
For the segmentation of the image a threshold value is used ranging from $0$ to $255$, discriminating between PPy (dark areas) and Si (bright areas). The predicted curves for the macroscopic Young\textsf{\char18}s modulus as function of the greyscale threshold are shown in Fig.~4 for the empty and PPy filled pSi membrane. The results are computed for orientations of the Si $<$110$>$ crystallographic direction at angles of $0\mathrm{^{\circ}}$ as well as $45\mathrm{^{\circ}}$ relative to the $x$-axis of the model.
	For the image with a threshold of 123, the predicted Young\textsf{\char18}s modulus of the empty pSi membrane under uniaxial tension orthogonal to the pore axis fits the measured macroscopic Young\textsf{\char18}s modulus $E=10\,\mathrm{GPa}$, a value derived from dynamical mechanical analysis (DMA), see sample fabrication.
	At $E=10\,\mathrm{GPa}$ (greyscale threshold = $123$), the anisotropy of Si is negligible. Hence, the structure of the pSi network dominates the macroscopic stiffness of the material. The upper inset in Fig.~4 presents the resulting microstructure of the pSi network (white) and pores (black), being highly irregular. For macroscopic tension in y-direction, the stress paths through the pSi network appear in green in the von Mises stress distribution in the lower inset whereas unstressed areas appear in blue. A comparison of both images reveals that only about $70\,\mathrm{\%}$ of the mass of the pSi network is contributing to the load transfer from the bottom to the top. Along with the progressive drop of the Young\textsf{\char18}s modulus of pSi in Fig.~4 towards a threshold of $138$, this indicates that the network is close to its percolation threshold.\\
	Towards the minimum and maximum greyscale values of $0$ and $255$, the curves approach the bulk Si and PPy properties, respectively. 
	\begin{figure}[]
		\centering
		\includegraphics[]{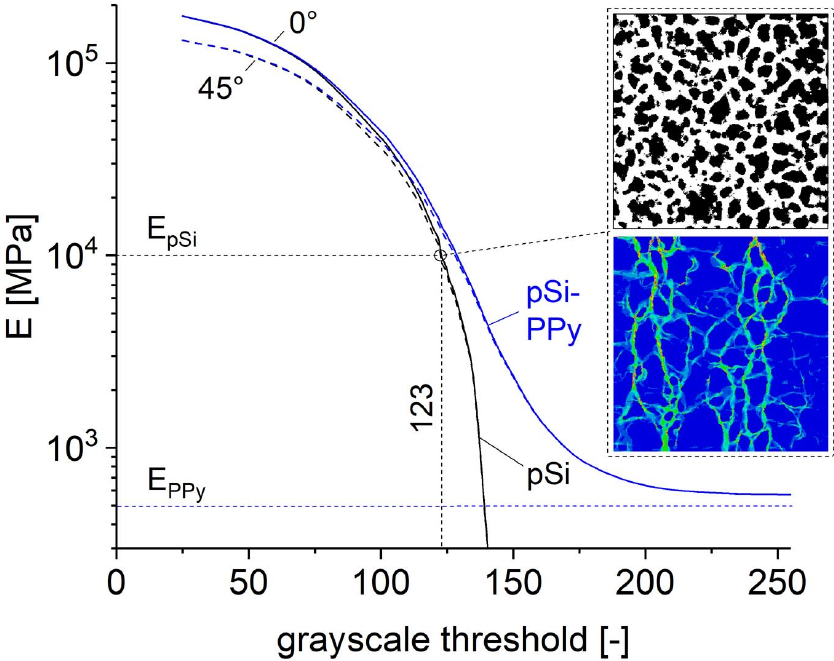}
		\caption{\textbf{Young\textsf{\char18}s modulus of the empty and PPy filled pSi membrane.} Values are predicted as function of the greyscale threshold value. The black curve corresponds to the empty pSi membrane, the blue curve is predicted for the PPy filled pSi membrane. Calibration of the pSi membrane to the measured macroscopic Young\textsf{\char18}s modulus of $E=10\,\mathrm{GPa}$ yields a greyscale threshold of $123$.}
		\label{Fig_EModul}
	\end{figure}
	At the greyscale threshold 123, calibrated with the macroscopic pSi membranes Young\textsf{\char18}s modulus, the simulations show a difference in the PPy filled and empty pSi membrane modulus, which is  $E_\mathrm{pSi-PPy}-E_\mathrm{pSi}\approx4\,\mathrm{GPa}$. The magnification of the effect by a factor of 8 relative to the Young\textsf{\char18}s modulus of PPy results from the improved connectivity of the pSi network, i.e. the PPy filling acts as a glue that enables a transfer of load through additional Si-walls, which do not contribute to load-bearing in the empty pSi membrane.\\
With the help of a pSi-PPy composite model, the swelling of the PPy in the pores is determined from the macroscopically measured expansion $\varepsilon= 0.05\,\%$ of the PPy filled pSi membrane, as measured from Fig.~2C, yielding the stress free swelling of the PPy $\varepsilon_\mathrm{swell,PPy}= 0.77\,\%$. Simulation results for the PPy filled pSi membrane at maximum swelling strain of the PPy are shown in Fig.~5. Recently, a value of $\psi = 0.17\,\mathrm{mm^3}/\mathrm{C}$ for the charge-strain coupling parameter of a PPy film that is clamped to a substrate was reported \cite{Roschning2019}. The volumetric charge of $0.04\,\mathrm{C}/\mathrm{mm^3}$ that is incorporated in our sample during the linear voltage increase from $0.4\,\mathrm{V}$ to $0.9\,\mathrm{V}$ would thus strain a PPy film by the value of $\varepsilon = 0.68\,\%$. This value is close to our result of $\varepsilon_\mathrm{swell,PPy}=0.77\,\%$ even though the PPy is not in the shape of a film but incorporated into pSi.\\
	Visible stresses in the pSi network, shown in  Fig.~5A as von Mises stress, range from $50$ to $100\,\mathrm{MPa}$. In contrast to uniaxial tension the swelling of the PPy leads to a more isotropic and uniform stress distribution in the Si walls, such that $83\,\%$ of the wall material is effectively strained.
	\begin{figure*}[]
		\centering
		\includegraphics[width=1\textwidth]{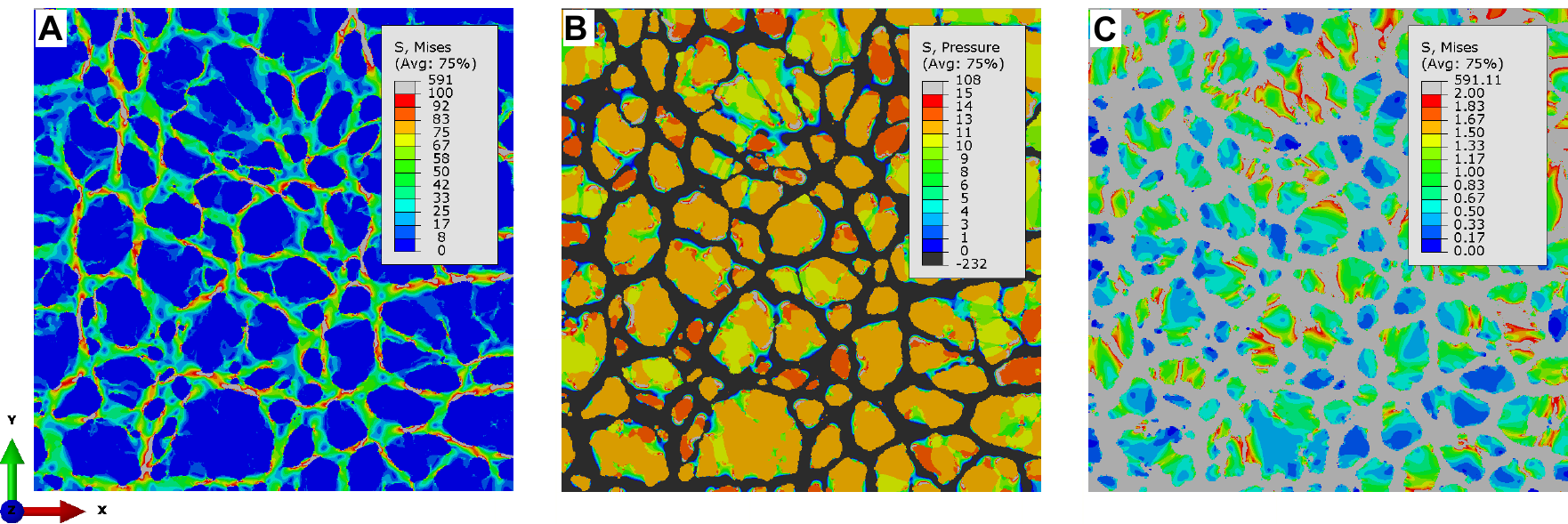}
		\caption{\textbf{Micromechanical analysis of electrochemical actuation.} Results of numerical simulations at maximum swelling strain:(\textbf{A}) von Mises stress distribution in the pSi walls, (\textbf{B}) pressure distribution in the PPy infiltrated pores, (\textbf{C}) von Mises stress distribution in the pores -- red colored areas exceed the $2\,\mathrm{MPa}$ yield stress of polypyrrole. }
		\label{Fig_Micromechanics}
	\end{figure*}
	From Fig.~5B can be seen that smaller pores tend to show higher pressures up to $15\,\mathrm{MPa}$, while larger pores, which may consist of a number of clusters of smaller pores and mechanically inactive Si walls, show pressures in the range from $8$ to $12\,\mathrm{MPa}$.
	Reducing the upper limit of the von Mises stress in the contour plot shown in Fig.~5C to the yield strength of $2\,\mathrm{MPa}$ for PPy as determined by Gnegel et al. \cite{Gnegel2019} reveals that, despite the high pressure in the pores, the polymer reaches the critical stress at which plastic flow occurs. Therefore, the numerical simulations support plasticity as possible reason for the observed differences in the actuation versus charging kinetics.\\ 
	In classic all-solid state piezoelectric systems a mechanical pressure ("Piezo", derived from the Greek $\pi\iota\acute{\varepsilon}\zeta\varepsilon\iota\nu$ (\textit{piezein}), which means to squeeze or press) couples to a macroscopic electrical polarization and thus to an electrical field via a deformation and thus movements of charged building blocks sitting on a crystalline lattice without inversion symmetry. This is not the case for our system. Here, an internal mechanical swelling pressure in the aqueous electrolyte/polypyrrole system builds up by movements of counter-ions into pore space as a function of an electrical potential applied to the entire porous medium with respect to the electrolyte. Thus a capacitive electrical charging of the entire volume of the porous medium is coupled to a deformation via an charge movement-induced pressure. \\
	How does the electrochemical actuator functionality of our composite compare with state of the art piezoelectric actuation? The piezoelectric coefficient of arguably the best performing commercially available lead- and lead-free piezoelectric ceramics (lead zirconate titanate (PZT: $(\mathrm{Pb}_{0.85}$$\mathrm{Ba}_{0.15})_{0.9925}$ $\mathrm{La}_{0.005}$\-$(\mathrm{Zr}_{0.52} \mathrm{Ti}_{0.48})\mathrm{O}_{3}$) and BZT-50BCT ($\mathrm{Ba}(\mathrm{Ti}_{0.8}\mathrm{Zr}_{0.2})\mathrm{O}_{3}$-$(\mathrm{Ba}_{0.7}\mathrm{Ca}_{0.3})\mathrm{TiO}_{3}$)), amount to $d_{33} = 410\,\mathrm{pCN^{-1}}$ and $620\,\mathrm{pCN^{-1}}$ \cite{Saito2004,Liu2009b}, respectively. Thus, to achieve a strain amplitude of $\varepsilon = 0.05\,\%$ for the same sample dimensions ($l=0.626\,\mathrm{mm}$) as studied here a potential of $763\,\mathrm{V}$ or $504\,\mathrm{V}$ would have to be applied, see also Fig.~3C. This is $840$ and $560$ times higher than for our lead-free, biocompatible pSi-PPy material evidencing the exceptional actuation performance of our hybrid system.\\ 
	Unfortunately, our step-coulombmetry experiments, discussed above, indicate that the functional kinetics is limited by the diffusion and drift-dynamics of the counter-ions in pore space. This results in long switching times, on the scale of seconds, much worse than achievable in piezoelectric systems, where switching times on the ms scale are easily reached. Here, a hierarchical pore structure encompassing nanopores with large actuation performance connected by macropores with fast bulk-like ion transport, similarly as often established in biological pore or capillary networks to optimize transport could lead to a substantial increase in the dynamic functional performance \cite{Zheng2017}. 
	Given the reversibility of both the counter-ion movement in the nanopores and of the mechanical response of the matrix, the electrochemomechanical coupling should also result in an incorporation or expulsion of ions and thus in an electrical voltage between porous medium and electrolyte, when an external pressure is applied to the porous medium in an open-circuit configuration. This so-called "piezoionic effect" is still controversially debated in the literature, being attributed either to stress gradient induced ion motion, to Donnan potentials arising at the polypyrrole/electrolyte interface or to the superposition of both \cite{Woehling2019}. However, for this reverse, sensoric function, we expect the coupling coefficient to be particularly small, since it should be inversely proportional to the large charge-strain coupling parameter $A^{*}$ determined above. Thus, our composite cannot compete with standard piezoelectric materials in terms of electrical voltage generation. Nevertheless, it will be interesting to explore this possible sensoric material functionality in the future.\\ 
	A large electrochemical actuation integrated into a mainstream semiconductor along with the huge varieties of structuring, surface modification, and functional integration of porous silicon \cite{Sailor2011,Canham2015} establishes particularly versatile and sustainable pathways to couple in an adaptive manner electronics with electrochemical energy storage\cite{Westover2014}, actuorics\cite{Baek2011,Gor2015,Huber2015}, fluidics\cite{Gruener2008,Vincent2014}, sensorics \cite{Lauerhaas1993,Lin1997} and photonics \cite{Atabaki2018} in aqueous electrolytic media. It thus extends the previous approaches to combine classic piezoelectric actuator materials, e.g. by thin film on-silicon coatings and epitaxial heterostructures to achieve on-chip actuorics, most prominently in the realm of nano- and micromechanical systems (NEMS/MEMS) \cite{Ho2008,Baek2011}. Similarly as established in nature in many biological composites \cite{Eder2018} the remarkably good functionality of the material results from a multiscale combination of light elemental constituents, here H, C, N, O, Si, Cl. In particular no heavy-metals, most prominently Pb, as omnipresent in high-performance piezoceramics, are necessary for the functionality. Given the biocompatibility and biodegradability of both porous silicon \cite{Canham2015} and polypyrrole \cite{Wang2004} along with the exceptionally small operation voltages this integrated material system may be particularly suitable for biomedical actuoric functions. From a more general materials science perspective, our study demonstrates how the advent of self-organized porosity in solids in combination with self-assembly and functionalization on the single-pore scale allows one to bridge the gap between bottom-up and top-down approaches for the design of 3-D mechanical robust materials, a particular challenge for embedding functional nanocomposites in macroscale devices \cite{Begley2019}.\\
	\section*{Materials and Methods}
	\subsection*{Sample fabrication} The silicon single crystals are p-doped and have a resistivity of $0.01-0.02\,\mathrm{\Omega cm}$, an (100) orientation and a thickness of $525\,\mathrm{\mu m}$. One side has been polished by the supplier (Si-Mat silicon Materials GmbH). The wafer is contacted by a thin piece of aluminium foil and mounted into an electrochemical setup with an inner radius of $1.58\,\mathrm{cm}$. It is made of polytetrafluorethylen (PTFE) to safely handle HF. Initially, the native silicon dioxide layer on the surface of the silicon wafer is stripped by immersing the silicon in a $1\%$ aqueous HF solution for $30\,\mathrm{s}$. The cell is then filled by a volumetric $4:6$ mixture of HF ($48\,\%$, Merck Emsure) and Ethanol (absolute, Merck Emsure). A round platinum counter electrode (CE) is implemented above the silicon in the electrochemical cell so that the platinum is submerged in the electrolyte. A current is applied between the platinum counter electrode (cathode) and the silicon working electrode (anode), in order to etch pores into the silicon. The applied current amounts to $98\,\mathrm{mA}$, which is equivalent to a current density of $12.5\,\mathrm{mA/cm^{2}}$. It is applied for $100\,\mathrm{minutes}$. In a final step the current is increased to detach the porous silicon layer from the remaining bulk silicon by means of an electro polishing process. Therefore, the current is set to $2\,\mathrm{A}$ for $30\,\mathrm{s}$. The resulting detached porous layer is rinsed three times with deionized water ($18.2\,\mathrm{M \Omega}$) and left to dry in ambient conditions for $2\,\mathrm{hours}$. \\
	The part of the fabricated pSi membrane, that is intended to be filled by a PPy polymerization, has a size of $1.008\,\mathrm{cm^2}$.
	A thin gold layer is deposited onto the detached side of the membrane. It has a thickness of $20 \pm 0.5\,\mathrm{nm}$. As already described in the main text, the membrane is placed in an electrochemical cell and a platinum pseudo reference electrode (RE) as well as a platinum counter electrode (CE) are inserted into the cell, see Fig.~S1B. The platinum RE has a stable potential of $0.2\,\mathrm{V}$ versus an Ag/AgCl RE. The solution for the electro polymerization is then added, which contains $0.1\,\mathrm{M}$ pyrrole monomers (Sigma Aldrich) as well as $0.1\,\mathrm{M}$ lithium perchlorate ($\mathrm{LiClO}_4$) (Sigma Aldrich) in acetonitrile (LiChrosolv, Merck) as a solvent. The pyrrole monomer is distilled at $130\mathrm{^{\circ}C}$ at ambient conditions immediately before the polymerization process in order to obtain a pyrrole solution with a high monomer amount. The $\mathrm{LiClO}_4$ is dissolved by stirring for $5\,\mathrm{minutes}$. The solution is transferred to the electrochemical cell and allowed to imbibe into the pores for $10\,\mathrm{minutes}$. The constant current of $0.403\,\mathrm{mA}$, that is applied for the electro polymerization, corresponds to a current density of $400\,\mathrm{\mu A / cm^2}$ per sample area. The reference voltage is recorded in steps of $0.1\,\mathrm{s}$ during the process. Afterwards the sample is rinsed three times with deionized water and allowed to dry under ambient conditions for $2$ hours.\\
	The charge $C$ consumed in the electro polymerization can be related to the weight $W_\mathrm{theor.}$ of the polymerized pyrrole by 
	\begin{equation}
	W_\mathrm{theor.} = \frac{C \cdot M}{z \cdot \mathrm{F}} ,
	\end{equation} where the \textit{Faraday} constant $\mathrm{F}$ equals $96485.332\, \mathrm{C}/\mathrm{mol}$. $M$ gives the formula weight of PPy, obtained from the molar mass of a monomer, while including incorporated $\mathrm{ClO}_4^{-}$-counterions and amounts to $96.78\,\mathrm{g}/\,\mathrm{mol}$ \cite{Harraz2006}. $z$ is the number of electrons used in the process to polymerize two monomers and is estimated by Diaz \textit{et al.} to $2.25$ \cite{Diaz1981}. For the electro polymerization presented in Fig.~1B of the main text $C$ amounts to $7.605\,\mathrm{C}$ which results in a theoretical weight of $W_\mathrm{theor.} = 0.00339\,\mathrm{g}$. The weight of the whole sample increases from $0.01005\,\mathrm{g}$, measured before the polymerization, to $0.01350\,\mathrm{g}$ afterwards. Hence, the actual weight increase is $W_\mathrm{meas.}=0.00345 \pm 0.00001\,\mathrm{g}$ and coincides well with the theoretical expectation.\\
	SEM micrographs with EDX are recorded to verify if the PPy is distributed equally along the whole thickness of the pSi membrane. The nitrogen signal in the EDX data is homogenous over the sample profile, which suggests a homogenous filling of the membrane over its thickness $t$. SEM micrographs are recorded with a Zeiss Supra 55 VP electron microscope at a potential of $5\,\mathrm{kV}$. TEM micrographs are recorded with a FEI Talos F200X.\\
	The sample is electrically contacted at the bottom by connecting a gold wire to the deposited gold layer. Subsequently, the sample is mounted. Therefore, the bottom of the sample is encased in epoxy for approximately $1\,\mathrm{cm}$. The sample is then placed in a glass beaker and installed in the dilatometer. It can then be immersed in $\mathrm{HClO}_4$. The sample that is in contact with the acid has a volume of $V_\mathrm{sample} = 0.186\,\mathrm{mm^3}$ and the volume of the pore space, using the porosity $\Phi$, amounts to $V_\mathrm{pores}=0.093\,\mathrm{mm^3}$.\\
	The electrochemical measurements are conducted in $1\,\mathrm{M}$ perchloric acid. The acid is prepared from $70\%$ $\mathrm{HClO}_4$ (Suprapure, Merck) and deionized water. The measurements are performed with a potentiostat equipped with a linear scan generator (Metrohm-Autolab PGSTAT 30) in a three electrode setup. A reversible hydrogen electrode (RHE) (Gaskatel HydroFlex) serves as the reference electrode and a carbon fabric as the counter electrode.\\
	The dilatometry measurements are conducted with a Linseis L75 dilatometer. Initially, the quartz pushrod of the dilatometer is carefully lowered onto the membrane with a static force of $0.4\,\mathrm{N}$ to install the sample into the dilatometer. It is effectively able to detect changes in the membrane length $l$ with a resolution of $12.5\,\mathrm{nm}$.\\
	The measurements to determine the macroscopic Young\textsf{\char18}s modulus of an empty pSi membrane are performed in a dynamical mechanical analysis (Netzsch DMA 242 C) with a three point bending setup. Thereby, a pSi membrane with a thickness of $100\,\mathrm{\mu m}$ is positioned on two contacts on the side while a pushrod applies a sinusoidal force in the middle of the sample until a certain strain amplitude is reached. From the force and the strain a Young\textsf{\char18}s modulus can be derived. The Young\textsf{\char18}s modulus of $10\,\mathrm{GPa}$ presented in this work has been found by averaging over many load and unload cycles as well as  upending the pSi membrane and repeating the measurements.\\ 
	%
	%	
	%________________________________________________________________
	%
	%
	%Micromechanical analysis.
	%
	%
	%________________________________________________________________
	\subsection*{Micromechanical analysis} The 2D FEM model of the pSi-PPy membrane is set up with the same discretization as the transmission electron micrograph, which is a \textit{.tiff} image consisting of $512\times512\,\mathrm{pixels}$. Assuming that the microstructure is constant through the thickness of the pSi membrane, each pixel of the image is represented by a 4-node quadrilateral plain strain element CPE4 in ABAQUS \cite{Abaqus2014}. The properties of each element are chosen according to the corresponding greyscale value of the image. For values below the greyscale threshold, the element is part of the pore space, for values equal or above the threshold, the element is added to the element set representing the Si network. The resulting model consists of $262144$ equally sized quadrilateral four-node elements organized in two element sets for assigning the Si and pore material properties. 
	Orthotropic elasticity is used to model the pSi membrane in the frame of a reference of a standard $(100)$ silicon wafer, which is $[110]$, $[110]$, $[001]$ with $E_1=E_2=169\,\mathrm{GPa}$, $E_3=130\,\mathrm{GPa}$, $\nu_{23}=0.36$, $\nu_{31}=0.28$, $\nu_{12}=0.064$, and $G_{23}=G_{32}=79.6\,\mathrm{GPa}$, $G_{12}=50.9\,\mathrm{GPa}$ \cite{Hopcroft2010}. Simulations are carried out for two orientations with $[110]$ in $x$-direction of the model ($0\mathrm{^{\circ}}$) and $[110]$ oriented $45\mathrm{^{\circ}}$ relative to the $x$-direction of the model ($45\mathrm{^{\circ}}$). For this rotation, the $z$-axis and $3$-direction $[001]$ serve as common rotation axis for the Si crystal. 
	For the simulation of the pure pSi membrane, the element set for the pores is assigned with isotropic elasticity $E_\mathrm{p}=1\,\mathrm{MPa}$, $\nu_\mathrm{p}=0.35$, i.e. the pore space has a negligible but non-zero stiffness. With this, all elements in the model are properly connected and the simulation is numerically stable. For simulations with a PPy filling inside the pore space, the elasticity parameters of the pore elements are set to $E_\mathrm{p}= E_\mathrm{PPy}=500\,\mathrm{MPa}$, $\nu_\mathrm{p}=\nu_\mathrm{PPy}=0.35$ \cite{Roschning2019}. Swelling of PPy is implemented by assigning a thermal expansion of $\alpha_\mathrm{PPy}=1\,\mathrm{K^{-1}}$ such that $\varepsilon_\mathrm{swell,PPy}=\varepsilon_\mathrm{th,PPy}=\alpha\Delta T$, where temperature difference  $\Delta T$ serves as an adjustable parameter. To model the effect of swelling as the sole effect of PPy, the thermal expansion for Si is set to $\alpha_{Si}=0\,\mathrm{K^{-1}}$. 
	For all simulations in ABAQUS, the boundaries are forced to remain plane using *EQUATION. Displacement boundary conditions are applied to the bottom and left boundaries such that the nodes do not move normal to the boundaries. The displacements of the top and right boundaries are controlled via dummy nodes included in the *EQUATION constraints. For macroscopic tension, the dummy node at the top boundary is displaced in positive $y$-direction. The $x$-displacement of the dummy node at the right boundary can be used to determine the Poisson\textsf{\char18}s ratio. For the swelling simulation, the displacements of both dummy nodes at the top and right boundaries are left free to move normal to the respective boundary.\\
	%
	%
	%Using this mechanism, the pSi membrane can be used as highly sensitive sensor material for the measurement of mechanical properties of polymers infiltrated and cross-linked in the nanopores.\\
	%
	%\noindent \textbf{Methods: Image processing in Fiji.} 
	%
	Further processing and analysis of the binarized image are done in several steps in the open-source program Fiji \cite{Schindelin2012}. 
	As the skeletonization is very sensitive to surface irregularities, the image is smoothed and again binarized. The smoothing process replaces each pixel with the average of its $3\times3$ neighborhood. 
	During this process, the volume fraction is increased by negligible $0.001$. The skeleton is constructed with the surface-thinning algorithm by Lee et al. \cite{Lee1994}, which is implemented as the process Skeletonize3D in the plugin BoneJ \cite{Doube2010} in Fiji. It converts an image object into its one-voxel-wide inner skeleton centerline. 
	If no surface smoothing is done prior to this, the resulting skeleton consists out of 2633 branches compared to 584, and 1461 junctions compared to 342.\\
	For the thickness measurement of the pSi walls, the Quasi-Euclidean 2D option of the Distance Transformation in the MorphoLibJ plugin \cite{Legland2016} is used. 
	The resulting Euclidean distance transform (EDT) image is multiplied with the skeleton image. As the resulting EDT values correspond to half of the wall thickness, the EDT values along the skeleton are multiplied by two. The same process was introduced by Badwe et al. \cite{Badwe2017} and Stuckner et al. \cite{Stuckner2017}.
	Besides the average wall thickness, the non-local values can be analyzed, as described in Richert and Huber 2018 \cite{Richert2018}.\\
	The average pore size of the smoothed and binarized image is computed using the thickness calculation \cite{Dougherty2007} in the plugin BoneJ \cite{Doube2010}. It constructs the diameter information at each voxel with the Biggest Sphere algorithm by Hildebrand et al. \cite{Hildebrand1997}. As this algorithm attributes one value to each pore space voxel, the average is volume-based instead of skeleton-based as the EDT approach, which makes it more suitable for the pore space.\\
	%
	%\noindent \textbf{Methods: Comparison with Grosman et al.}\\
	Grosman et al. \cite{Grosman2015} analyzed the macroscopic Young\textsf{\char18}s modulus of pSi assuming that the Si walls are organized in a honeycomb structure, using the in-plane elastic constants from \cite{Gibson1997}. From the discrepancy of the measurements with the calculated values, the authors concluded that  the Young\textsf{\char18}s modulus of the Si walls is $5$ times smaller than that of bulk silicon and explained this with the finite-size effect that can decrease the effective Young\textsf{\char18}s modulus.\\
	However, due to the strong irregularity of the network and the uncertainty concerning the threshold, it is not trivial to characterize the geometry of the pSi network. Here, the calibration of the greyscale threshold via the macroscopic Young\textsf{\char18}s modulus provides a very sensitive measure for the segmentation of the image into Si and pore space, which can then be analyzed using Fiji \cite{Schindelin2012}. The segmentation determines the fraction of Si in the membrane to $0.508$, which is in excellent agreement with the measured porosity $\Phi = 50\,\mathrm{\%}$. Using the Euler Distance Transformation \cite{Legland2016} and image multiplication with the skeleton \cite{Lee1994,Doube2010}, the average thickness of the Si walls is determined to $t_\mathrm{Si}=9.95\,\mathrm{nm}\approx 10\,\mathrm{nm}$ with a standard deviation of $4\,\mathrm{nm}$. These values are used in the following for a comparison with the predicted properties based on a honeycomb structure.\\
	For completeness, the average diameter and standard deviation of the pores is measured with the volume-based Biggest Sphere algorithm \cite{Hildebrand1997}, yielding $18.9\,\mathrm{nm}$ and $6.\,\mathrm{nm}$, respectively.
	The junctions between walls are on average a 3-fold connected, which equals the connectivity in a honeycomb structure. The average wall length in the fully-interconnected network is determined as $19.1\,\mathrm{nm}$. Besides the average values, the network can furthermore be analyzed non-locally, which is here achieved by object-oriented programming as in the work of Richert et al. \cite{Richert2018}. The average thickness of a wall along its axis is not constant, but can be classified as a slightly concave profile. At the junctions the network is on average $13.0\,\mathrm{nm}$ thick, and at the thinnest part of a wall on average $8.6\,\mathrm{nm}$ thick.\\
	For an average wall size of $t=10\,\mathrm{nm}$ and a porosity of $50\,\mathrm{\%}$, the computed wall length of a honeycomb structure is $l=20\,\mathrm{nm}$. According to \cite{Gibson1982}, the effective Young\textsf{\char18}s modulus under uniaxial tension is given by
	\begin{equation}
	E = \frac{4}{\sqrt{3}}\left(\frac{t}{l}\right)^3 E_{Si}\;.
	\label{Gibson}
	\end{equation}
	Inserting $t/l=0.5$ and $E=10\,\mathrm{GPa}$ yields an estimate for $E_\mathrm{Si}$ of only $35\,\mathrm{GPa}$, which is comparable to the value reported in \cite{Grosman2015}. However, our analysis shows that 30\% of the Si material carries little or no load, see inset in Fig.~4. Furthermore, considering the measured standard deviation of the wall thickness which is almost $50\,\mathrm{\%}$ of the average wall thickness, the idealization with a regular honeycomb structure is highly questionable. For instance, reducing the $t/l$ ratio in Eq. (2) to $t/l=0.31$ already yields a Young\textsf{\char18}s modulus of $E_\mathrm{Si}=145\,\mathrm{GPa}$, which corresponds to the average bulk Young\textsf{\char18}s of Si. Therefore, the high sensitivity of the macroscopic stiffness with respect to the $t/l$ ratio along with the drop in structural stiffness caused by partially deactivated Si walls (see also \cite{Huber2018} for 3D networks) underlines the importance of a careful characterization and translation of the pSi membrane into a micromechanical model that reflects the real irregular microstructure, as presented in the upper insert of Fig.~4.\\
	\textbf{Acknowledgements:}\\ 
	We acknowledge fruitful discussions with, and a critical reading by J\"org Weissm\"uller (Hamburg University of Technology). Moreover, we acknowledge support by the Hamburg Centre for Integrated Multiscale Materials Systems (CIMMS).\\
	\textbf{Funding:}\\
	This work was supported by the Deutsche Forschungsgemeinschaft (DFG) within the Collaborative Research Initiative SFB 986 "Tailor-Made Multi-Scale Materials Systems" Projectnumber 192346071.\\
	\textbf{Author contributions:}\\
	M.B. and P.H. conceived the experiments. M.B. and G.D. performed the material synthesis and the actuator measurements. P.L. and T.K. contributed to the design of the actuation experiments. The transmission electron micrographs were taken by M.B. and T.K. C.R. and N.H. performed the micromechanical analysis. M.B., C.R., N.H. and P.H. wrote the manuscript. All other authors proofread the manuscript.\\ 
	\textbf{Competing interests:}\\
	The authors declare that they have no competing interests.\\
	\textbf{Data and materials availability:}\\
	All data needed to evaluate the conclusions in the paper are present in the paper and/or the Supplementary Materials. The raw data of the electrochemical and electroactuation experiments, a complementary transmission electron micrograph as well as the micromechanical analysis are available at TORE (https://tore.tuhh.de/), the Open Research Repository of Hamburg University of Technology at the doi: https://doi.org/10.15480/336.2753.\\
\bibliographystyle{ScienceAdvances}
\end{document}